\documentstyle[amsmath,amssymb,graphicx]{article}

\def\be{\begin{eqnarray}}
\def\ee{\end{eqnarray}}

\def\l[{\phantom.[}

\hoffset= - 1.0in         

\begin{document}

\title{\vspace{.1cm}{\Large {\bf  Eigenvalue conjecture and colored Alexander polynomials}\vspace{.2cm}}
\author{
{\bf A.Mironov$^{a,b,c,d}$}\footnote{mironov@lpi.ru; mironov@itep.ru}\ \ and
\ {\bf A.Morozov$^{b,c,d}$}\thanks{morozov@itep.ru}}
\date{ }
}

\maketitle

\vspace{-5.5cm}

\begin{center}
\hfill FIAN/TD-23/16\\
\hfill IITP/TH-17/16\\
\hfill ITEP/TH-23/16
\end{center}

\vspace{3.3cm}

\begin{center}
$^a$ {\small {\it Lebedev Physics Institute, Moscow 119991, Russia}}\\
$^b$ {\small {\it ITEP, Moscow 117218, Russia}}\\
$^c$ {\small {\it Institute for Information Transmission Problems, Moscow 127994, Russia}}\\
$^d$ {\small {\it National Research Nuclear University MEPhI, Moscow 115409, Russia }}

\end{center}

\vspace{.5cm}

\begin{abstract}
We connect two important conjectures in the theory of knot polynomials.
The first one is the property $Al_R(q) = Al_{[1]}(q^{|R|})$ for all single hook
Young diagrams $R$, which is known to hold for all knots.
The second conjecture claims that all the mixing matrices $U_{i}$ in the relation
${\cal R}_i = U_i{\cal R}_1U_i^{-1}$ between the $i$-th and the first generators
${\cal R}_i$ of the braid group are universally expressible through the eigenvalues
of ${\cal R}_1$.
Since the above property of Alexander polynomials is very well tested,
this relation provides a new support to the eigenvalue conjecture,
especially for $i>2$, when its direct check by evaluation of the Racah matrices and their
convolutions is technically difficult.
\end{abstract}

\vspace{1.5cm}

An undisputable advantage of {\it knot theory} from the point of view of {\it representation theory}
is that the former provides a set of quantities, which adequately capture and reveal the basic hidden
properties of the latter.
These quantities, knot polynomials \cite{knotpols} are the most natural in
quantum field theory (QFT) realization of knot theory:
they are just the Wilson loop averages in the topological Chern-Simons model \cite{CS},
which is one of the simplest in the family of Yang-Mills theories.
The power of knot polynomial (QFT) methods in knot and representation theories
is an impressive manifestation of the effectiveness of string theory approach to
mathematical problems, especially when their calculational (algebraic) aspects are concerned.

This letter is an attempt to "explain" the amusing property of Alexander polynomials, the specializations of the (colored) HOMFLY polynomials at $A=1$,
\be
Al_R^{\cal K}(q) := H_R^{\cal K}(A=1,q)
\ee
The property is that
\be
Al_R^{\cal K}(q) = Al_{_\Box}^{\cal K}\left(q^{|R|}\right)
\label{Alred}
\ee
for all knots ${\cal K}$ and all single hook Young diagrams $R=[r,1^s]$
of size $|R|=r+s$.
This phenomenon was "experimentally" observed in \cite{IMMMfe} (and later discussed in \cite{Zhu})
as a "dual" to the property of {\it special} polynomials \cite{DMMSS} (discussed in \cite{speproof,Zhu} and generalized to superpolynomials in \cite{Anton}), arising from the HOMFLY polynomial at $q=1$,
\be
H_R^{\cal K}(A,q=1) = \left(H_{_\Box}^{\cal K}(A,q=1)\right)^{|R|}
\ee
which is in fact valid for {\it all} Young diagrams $R$ (not obligatory single hook),
and was actually proved rather fast \cite{speproof}.
Relation (\ref{Alred}), however, got less attention and still remains a mystery.
In the present text, we reduce it to another "experimental" discovery,
{\it the eigenvalue conjecture} of \cite{IMMMev}, moreover, (\ref{Alred})
follows from its {\it stronger} version, applicable to arbitrary number $m$ of strands
in the braid (the week form considered in most detail in \cite{IMMMev} concerned only
$m=3$ and the ordinary Racah matrices).

\bigskip

Namely, we decompose the explanation of (\ref{Alred}) into five steps:

\bigskip

$\bullet$ Realize the knot ${\cal K}$ as a closure of an $m$-strand braid;
a lot of such realizations is possible for any ${\cal K}$,
equivalence of the HOMFLY polynomials for different choices is guaranteed
by invariance under the Reidemeister moves.
Then (see \cite{RT,DMMSS},\cite{RTmodf}-\cite{RTmodl})
\be
H_R^{\cal K}(A,q) = \sum_{Q \vdash m|R|} C_{RQ}(q) \cdot\frac{\chi_Q^*}{\chi_R^*}(A,q)
\label{decoHOMFLY}
\ee
where all the dependence on $A$ is localized in the quantum dimensions ($q$-graded traces)
$\chi^*$.

\bigskip

$\bullet$ The quantum dimensions
are given by the (hook) product  formula over all
boxes of the Young diagram:
\be
\chi_Q^* = \prod_{\square\in Q} \frac{\{Aq^{l_{_\square}'-a_{_\square}'}\}}
{\{q^{l_{_\square}+a_{_\square}+1}\}}
\ee
where $\{x\}=x-x^{-1}$, while $l,a,l',a'$ are the lengths of legs, arms, co-legs and co-arms respectively:

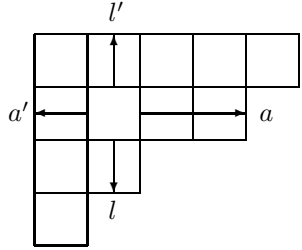
\begin{picture}(120,105)(-150,13)
\put(20,100){\line(1,0){100}}
\put(20,80){\line(1,0){100}}
\put(20,60){\line(1,0){80}}
\put(20,40){\line(1,0){40}}
\put(20,20){\line(1,0){20}}
\put(20,100){\line(0,-1){80}}
\put(40,100){\line(0,-1){80}}
\put(60,100){\line(0,-1){60}}
\put(80,100){\line(0,-1){40}}
\put(100,100){\line(0,-1){40}}
\put(120,100){\line(0,-1){20}}

\put(40,70){\vector(-1,0){20}}
\put(50,80){\vector(0,1){20}}
\put(60,70){\vector(1,0){40}}
\put(50,60){\vector(0,-1){20}}

\put(10,67){\mbox{$a'$}}
\put(48,105){\mbox{$l'$}}
\put(105,67){\mbox{$a$}}
\put(48,30){\mbox{$l$}}
\end{picture}

\noindent
Contributing to (\ref{decoHOMFLY}) at $A=1$ are diagrams $Q$ with no $l=a$ factors
(which vanish at $A=1$).
One such factor is obligatory present for any $Q$, but it drops
away from the ratio $\chi^*_Q/\chi_R^*$.
For a single hook $R=[r,1^s]$, however, contributing $Q$ are also only
single hook.
Since their sizes are $|Q|=m|R| = m(r+s)$, and at the same time the numbers of rows
and columns are restricted by $mr$ and $m(s+1)$ respectively,
the set of  $Q$ contributing to (\ref{decoHOMFLY}) at $A=1$ consists just of $m$
single hook diagrams
\be
Q = [mr-k,1^{ms+k}], \ \ \ \ k = 0,\ldots,m-1
\label{Q1hook}
\ee
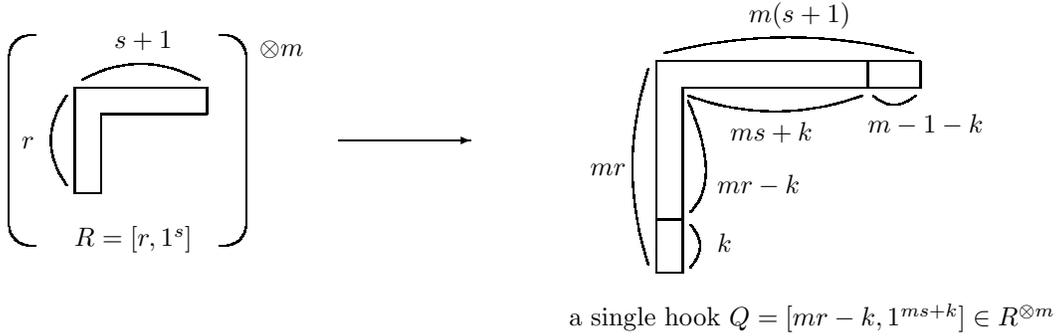
\begin{picture}(120,140)(-270,-14)
\put(20,100){\line(1,0){100}}
\put(30,90){\line(1,0){90}}
\put(20,20){\line(1,0){10}}
\put(20,40){\line(1,0){10}}
\put(20,100){\line(0,-1){80}}
\put(30,90){\line(0,-1){70}}
\put(120,100){\line(0,-1){10}}
\put(100,100){\line(0,-1){10}}
\qbezier(23,103)(70,115)(117,103)
\put(55,115){\mbox{$m(s+1)$}}
\qbezier(17,23)(5,60)(17,97)
\put(-5,57){\mbox{$mr$}}
\qbezier(33,22)(40,30)(33,38)
\put(43,28){\mbox{$k$}}
\qbezier(33,43)(45,55)(32,85)
\put(43,50){\mbox{$mr-k$}}
\qbezier(102,87)(110,80)(118,87)
\put(100,73){\mbox{$m-1-k$}}
\qbezier(32,87)(65,75)(98,87)
\put(48,70){\mbox{$ms+k$}}
\put(-200,90){\line(1,0){50}}
\put(-190,80){\line(1,0){40}}
\put(-200,50){\line(1,0){10}}
\put(-200,90){\line(0,-1){40}}
\put(-190,80){\line(0,-1){30}}
\put(-150,90){\line(0,-1){10}}
\qbezier(-197,93)(-175,105)(-153,93)
\put(-185,105){\mbox{$s+1$}}
\qbezier(-203,53)(-215,70)(-203,87)
\put(-220,67){\mbox{$r$}}
\put(-100,70){\vector(1,0){50}}
\put(-130,102){\mbox{$\otimes m$}}
\put(-135,100){\line(0,-1){60}}
\put(-225,100){\line(0,-1){60}}
\qbezier(-225,100)(-225,110)(-215,110)
\qbezier(-225,40)(-225,30)(-215,30)
\qbezier(-135,100)(-135,110)(-145,110)
\qbezier(-135,40)(-135,30)(-145,30)
\put(-200,30){\mbox{$R=[r,1^s]$}}
\put(-13,0){\mbox{a single hook $Q = [mr-k,1^{ms+k}]\in R^{\otimes m}$}}
\end{picture}

$\bullet$
If the knot ${\cal K}$ is realized
as a closure of the $m$-strand braid
$\Big(a_{1,1},\ldots,a_{1,m-1}|a_{2,1},\ldots a_{2,m-1}| \ldots\Big)$,
then the coefficients $C_{RQ}$ in (\ref{decoHOMFLY})
are actually equal to \cite{RTmodf}-\cite{RTmodl}
\be
C_{_{RQ}} = {\rm Tr}_{_{V_Q}} \Big(  {\cal R}_1^{a_{1,1}}\ldots \ {\cal R}_{m-1}^{a_{1,m-1}}
\,{\cal R}_1^{a_{2,1}}\ldots \ {\cal R}_{m-1}^{a_{2,m-1}}\, \ldots\Big)
.\ee
where the trace is taken over the space of intertwining operators (multiplicities), $R^{\otimes m} = \oplus_Q \, V_Q^{(m)}\otimes Q$
and the ${\cal R}$-matrix ${\cal R}_i$ standing at the intersection of $i$-th strand
with the $(i+1)$-th one is obtained by the conjugation
\be
{\cal R}_i = U_i{\cal R}_1U_i^{-1}
\ee
It is associated with the fact that the usual ${\cal R}$-matrix which acts in the space of product of two representations $R\otimes R$
if diagonalized, is proportional to unity in the space of the irreducible representation $Y$ that appears in the
decomposition of the {\it square} of ${R}$, $R\otimes R = \oplus_Y \ V^{(2)}_Y\otimes Y$:
\be\label{ev}
{\cal R}_i= \oplus_Y \ \epsilon_i q^{ \varkappa_{_Y} }\cdot I_{V_Y^{(2)}}
\ee
where $\epsilon_i = \pm 1$, depending on whether $Y$ belongs to the symmetric or antisymmetric square.

This is why one can work with ${\cal R}$ that acts already in the space of intertwining operators $V^{(2)}_Y$. One can diagonalize, say, ${\cal R}_1$ with acts on the first two strands. However, all matrices ${\cal R}_i$ can not be diagonalized at once.
For $m>2$ each multiplicity space $V_Q^{(2)}$,
arising in the $m$-th tensor power of $R$, contains descendants of different $Y$
from the second level, and while ${\cal R}_1$ remains diagonal,
the other ${\cal R}_i$ (and thus elementary building blocks of $U_i$ \cite{MMMII,AMMM}) after a proper ordering of columns and rows are block-diagonal matrices with blocks of the size $V_Q^{(2)}\otimes V_Q^{(2)}$ \cite{MMMII,AMMM,Anopaths}.

\bigskip

If both $R$ and $Q\in R^{\otimes m}$ are single hook diagrams, then $Y\in R^{\otimes 2}$,
lying on the path from $R$ to $Q$ in the representation graph \cite{Anopaths},
are also single hook, and there are just two of the present, those
with $k=0$ and $k=1$.
Up to a common $R$-dependent shift, the eigenvalues (\ref{ev}) are equal to
\be
\varkappa_{_Y} = {\rm shift} +  \sum_{(i,j)\in Y} (i-j) \ \stackrel{(\ref{Q1hook})}{=}
 {\rm shift} \pm \underbrace{(r+s)}_{|R|}
 \label{ev1h}
\ee
for  $Y=[2r,1^{2s}]$ and $Y=[2r-1,1^{2s+1}]$.
In other words, 
the difference between eigenvalues for representation/Young diagram $R$ and for the fundamental
representation $\Box$ is exactly in the $|R|$-th power of $q$, the same as in (\ref{Alred}) (this is since the common shift is irrelevant: it is the same in the both cases in the topological framing).

\bigskip

$\bullet$ It is a simple exercise to check that exactly the same is true for the ratios
of quantum dimensions
$\chi_Q^*/\chi_R^*$:
\be
\left.\frac{\chi^*_{[rm-k,1^{ms+k}]} }{\chi^*_{[r,1^s]}}\right|_{A=1} = \frac{[r+s]}{[m(r+s)]}
= \frac{\big[|R|\big]}{\big[m|R|\big]}
\ee
where the quantum numbers are defined as $[N]=\frac{\{q^N\}}{\{q\}} = \frac{q^N-q^{-N}}{q-q^{-1}}$.

Keeping all this in mind,
for the $m=2$-strand knots with odd $n$, one obtains from (\ref{decoHOMFLY})
\be
Al_{[r,1^s]}^{(n)} = \frac{[r+s]}{[2(r+s)]}\cdot\Big(q^{n(r+s)}+q^{-n(r+s)}\Big)
= \frac{q^{n(r+s)}+q^{-n(r+s)}}{q^{r+s} +q^{-(r+s)}}
= \frac{q^{n|R|}+q^{-n|R|}}{q^{|R|} +q^{-|R|}}
\ee
which for odd $n$ (i.e. for the 2-strand {\it knots} rather than links) is indeed a polynomial,
satisfying
(\ref{Alred}).

\bigskip

$\bullet$
For $m>2$ just the same reasoning would work with the only correction:
the mixing matrices $U$ emerge.
For (\ref{Alred}) to be true, it is {\it sufficient} if $U$-matrices depend on $R=[r,1^s]$
through $q^{r+s}=q^{|R|}$ only -- and this is exactly what follows from the
{\it eigenvalue conjecture} of \cite{IMMMev}.
The conjecture claims that the $V_Q\otimes V_Q$ block of $U$,
associated with representation $Q$, are made entirely from the {\it normalized}
eigenvalues $\epsilon_Y q^{\varkappa_Y}$
of the ${\cal R}$-matrix ${\cal R}_1$ for $Y$ on the path from $R$ to $Q$,
and (\ref{ev1h}) shows that in our single hook
case these are in turn made exactly from $q^{r+s}=q^{|R|}$
("normalized" means that the "shifts" in (\ref{ev1h}) can be neglected).

It is of course important in this argument that
the  paths in representation graph
from a single hook $R$ to a single hook $Q$
are restricted to single hook diagrams at all steps.
In result, the encountered diagrams and, thus, the
entire paths are labeled by parameters $k$, which are actually $R$-independent:
the {\it paths} and contributing eigenvalues are just the same for the fundamental representation
and for any other single hook $R$.

\bigskip

\framebox{
Thus, {\bf the eigenvalue conjecture implies (\ref{Alred})}.}

\bigskip

This is the main claim of the present letter.
However, in practice this statement is rather an evidence in favor of the eigenvalue
conjecture than of (\ref{Alred}).
This is because (\ref{Alred}) is very easy to check, once colored HOMFLY is known,
and recent advances in HOMFLY calculus \cite{pret}-\cite{recthomfly} provided quite a number of examples,
what allows one to consider (\ref{Alred}) rather well tested.
The situation with the eigenvalue conjecture is much worse:
it was quite difficult to check it for the matrix-sizes (dimensions of $W(Q)$) 2,3,4,5
even in the simplest case of $m=3$ strands.
For size $6$ it was validated very recently within the framework of the knot universality of \cite{univ}
and by application to advanced Racah calculus in \cite{MMMS21,MMMS31,MMMS22}.
This was an important step, because, beginning from the size $6$, the
eigenvalue conjecture does not immediately follow from Yang-Baxter relations only \cite{IMMMev,math},
still for knot calculus it works well.
Evidence for the eigenvalue hypotheses for higher number of strands $m>3$ is still
nearly negligible.
Mixing matrices are now not just Racah matrices but their convolutions \cite{gmmms2},
which are extremely difficult to calculate, and not much has been yet done since \cite{RTmodf}-\cite{RTmodl}.
What we provided in this text, is a claim that, if true for all $m$, the eigenvalue conjecture
would explain the well established (\ref{Alred}), and this is a new and reasonably
strong evidence.
It is far from proving anything, both because (\ref{Alred}) is not proved
and because the eigenvalue conjecture is sufficient, but not necessary for (\ref{Alred}) to hold.
Still, this new relation between the two conjectures should attract new attention to both of
them and hopefully lead to a considerably better understanding.

From this interpretation of (\ref{Alred}) it gets clear, what is so special about the
single hook diagrams.
For $R$ with $h$ hooks contributing to the sum (\ref{decoHOMFLY}) at $A=1$ will be also the
$h$-hook diagrams $Q$, which will be parameterized by $2h-1$ parameters instead of a single $k$.
Analysis of this situation is now possible and straightforward, however,
it is not a surprise that the answer is more sophisticated than (\ref{Alred}).


\section*{Acknowledgements}

This work was performed at the Institute for Information Transmission Problems with the financial
support of the Russian Science Foundation (Grant No.14-50-00150).

\end{document}